\documentclass[12pt]{iopart}
\usepackage{iopams}
\usepackage{graphicx}

\begin{document}

\title[Measurement of the Casimir force with a ferrule-top sensor]{Measurement of the Casimir force with a ferrule-top sensor}

\author{P.~Zuurbier, S.~de~Man, G.~Gruca, K.~Heeck, and D.~Iannuzzi}
\address{Department of Physics and Astronomy and LaserLaB, VU University Amsterdam, Amsterdam, The Netherlands}
\ead{iannuzzi@few.vu.nl}

\begin{abstract}
We present a Casimir force setup based on an all-optical ferrule-top sensor. We demonstrate that the instrument can be used to measure the gradient of the Casimir force between a gold coated sphere and a gold coated plate with results that are comparable to those achieved by similar atomic force microscope experiments. Thanks to the monolithic design of the force sensor (which does not require any optical triangulation readout) and to the absence of electronics on the sensing head, the instrument represents a significant step ahead for future studies of the Casimir effect under engineered conditions, where the intervening medium or the environmental conditions might be unsuitable for the use of more standard setups.
\end{abstract}

 \maketitle

\section{Introduction}

Long range surface interactions are of paramount importance in the design of Micro- and NanoElectroMechanical Systems (MEMS and NEMS), as they determine the minimum separation that two miniaturized mechanical pieces can reach before they snap to contact. It is thus not surprising that, over the last decade, an ever increasing number of groups has been drawing the attention of the scientific community to the potential relevance of the Casimir effect in nanotechnology~\cite{Casimir:1948p440,Capasso:2007p1738} and on what currently goes under the name of \textit{quantum fluctuations engineering} -- the possibility of tailoring the Casimir force with a suitable choice of the shape and material properties of the interacting objects and of the medium between them~\cite{Iannuzzi:2004p117,Lisanti:2005p3,Chen:2006p89,Chen:2007p15,Munday:2009p9,deMan:2009p99,Chan:2008p81,Torricelli:2010p1674}\footnote{Over the last 30 years, there has been a much more extensive activity focused on the investigation of the van der Waals interaction in the non-retarded limit, with particular emphasis to liquid environments, and there are important examples in which the investigation has been extended to the retarded part of the interaction. A complete review of that part of the literature is out of the scope of this paper. We refer the reader to \cite{Lee:2001p1810,parsegian} for more details.}. Driven by this trend, scientists have developed a wide variety of instruments that can assess different aspects of this interaction mechanism. Macroscopic setups~\cite{Lamoreaux:1997p31,Kim:2009p643,Masuda:2009p1887,Bressi:2002p1735} and micromachined torsional balances~\cite{Chan:2001p73,Chan:2001p36,Decca:2003p32} are typically optimized for utilization in vacuum or air, but would hardly work in liquids. Experiments in vacuum can be as well performed by means of custom made atomic force microscopes (AFMs)~\cite{Mohideen:1998p63,Jourdan:2009p1405}, which, after proper modifications, can be also used to measure the Casimir force in gaseous environments~\cite{deMan:2009p99} or in liquids~\cite{Munday:2009p9}. Because AFMs rely on optical triangulation, however, it is difficult to imagine a universal measuring head that can easily adapt to different environments, ranging, for example, from low temperature vacuum to room temperature liquids.

Earlier this year, our group proposed to overcome this issue by replacing the AFM head with an all-optical micromachined torsional force sensor that adapts well to both vacuum and critical environments~\cite{deMan:2010p1670}. The sensor is based on fiber-top technology~\cite{Iannuzzi:2006p97}. It consists of a mechanical rectangular beam carved out of the cleaved end of a standard single mode optical fiber. The beam is suspended a few microns above the rest of the fiber by means of two lateral torsional rods. The light coupled from the opposite end of the fiber allows one to measure the tilting angle of the rectangular beam and, therefore, the force that makes it tilt. Thanks to its monolithic design and to the absence of electronics on the sensing element, this micro-opto-mechanical balance can be in principle used in any environment without any change of the readout mechanics, optics, or electronics. Unfortunately, however, preliminary experiments show that, as soon as measurements are not carried out in vacuum, the sensor can only be used in static mode~\cite{deMan:2010p1670}. Dynamic modes, which are typically more sensitive, are in fact disturbed by spurious effects induced by the hydrodynamic force between the mechanical beam and the fiber below (a phenomenon that goes under the name of \textit{squeezed field air damping}~\cite{bao}). Furthermore, because the optical fiber is only 125 $\mu$m in diameter, fiber-top devices are typically fabricated with an expensive and time consuming technique (namely, Focused Ion Beam (FIB) milling~\cite{Deladi:2006p1739}). Fiber-top technology cannot thus be considered as a practical solution for systematic measurements, where, due to recurrent accidental damaging of the force sensor, one must rely on probes that can be easily replaced.

To overcome the fabrication issue of fiber-top devices, we recently introduced a novel approach that preserves the flexibility of fiber-top technology while reducing manufacturing costs and production time: the ferrule-top cantilever~\cite{Gruca:2010p1677}. To fabricate a ferrule-top cantilever, a standard single mode optical fiber is glued inside the bore hole of a much bigger pierced ferrule. The fiber and the ferrule are so well held together by the glue that they behave like a single mechanical piece. The ferruled fiber is thus equivalent to a very large single mode optical fiber that can now be milled in the form of a cantilever by means of more convenient techniques (e.g., laser ablation). Interestingly, because of the larger dimensions of the building block, the gap between the cantilever and the remaining part of the ferrule is typically much larger than in fiber-top devices. Ferrule-top cantilevers are thus supposed to suffer considerably less from the hydrodynamic problems than fiber-top sensors.

In this paper we present a ferrule-top force setup designed to measure the Casimir attraction between a sphere and a flat plate, and we demonstrate that one can indeed perform precise measurements of the Casimir force between a sphere and a plate kept in air with a dynamic detection scheme that does not induce any spurious effects.

\section{Experimental setup}

The experimental setup presented in this paper is designed to measure the Casimir force between a 200 $\mu$m diameter sphere and a plate as a function of separation in a distance range between, approximately, 50 nm and 200~nm.

\begin{figure}[t]
\begin{center}
\includegraphics[width=16cm]{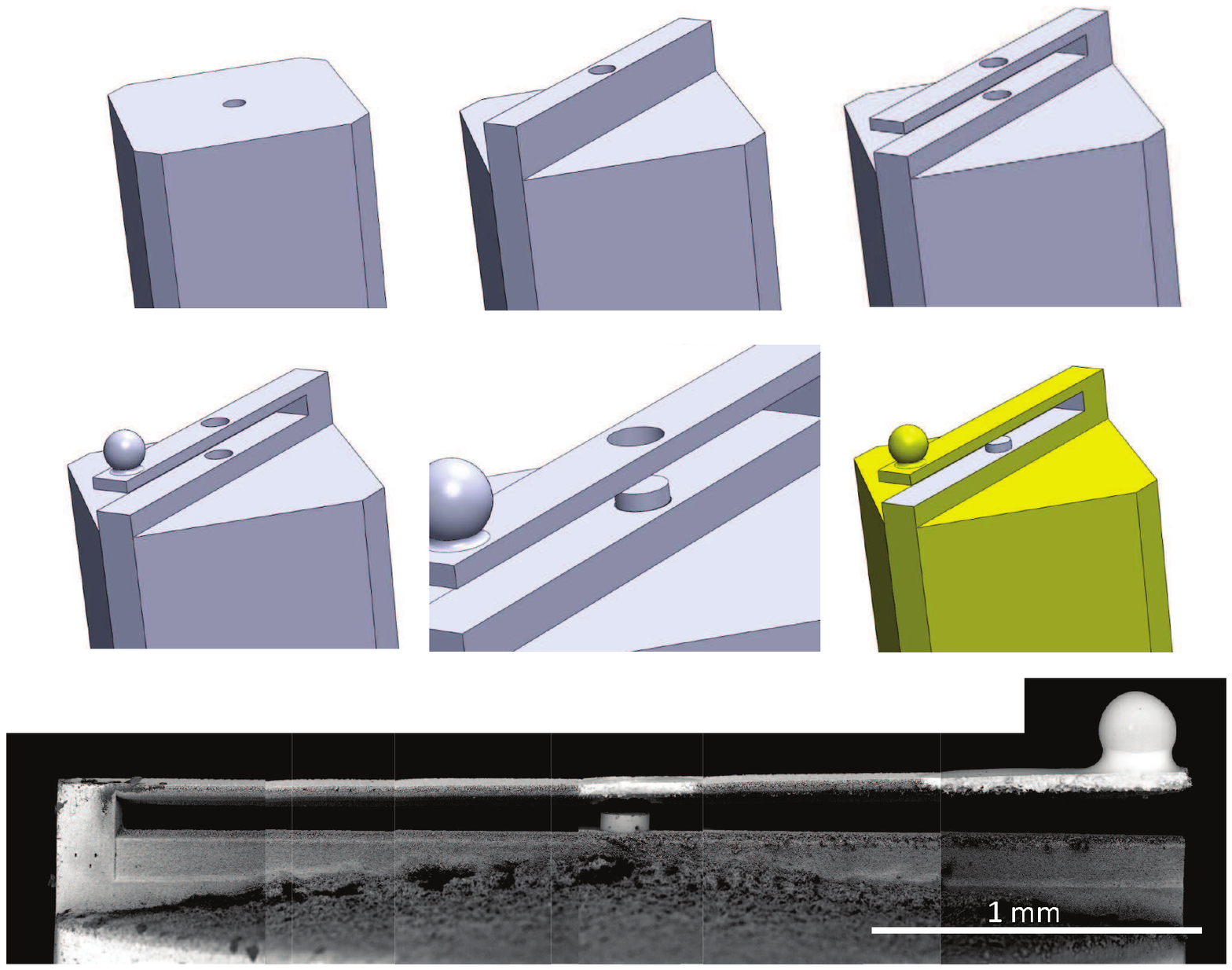}
\label{fig_fab}
\end{center}
\end{figure}

The force sensor is realized according to the scheme sketched in Fig.~\ref{fig_fab}. A pierced 2.5~mm $\times$ 2.5~mm $\times$ 7~mm rectangular ferrule, made out of borosilicate glass, is initially carved by means of laser ablation in the form of a cantilever that stretches over one of the diagonals of the edge of the ferrule. At the end of the milling process, a small amount of transparent epoxy is dropped and cured inside the 127 $\mu$m diameter hole left open at the center of the cantilever, while a standard single mode optical fiber is slid into the hole of the ferrule from the other side and glued with the cleaved end at approximately 100 $\mu$m from the bottom surface of the cantilever. A 200~$\mu$m diameter sphere is then attached to the top of the free hanging end of the sensor by means of a small droplet of UV curable epoxy. The sensor and the sphere are finally coated with a 5 nm thick Cr adhesion layer followed by a 200 nm thick Au film.

\begin{figure}[t]
\begin{center}
\includegraphics[width=16cm]{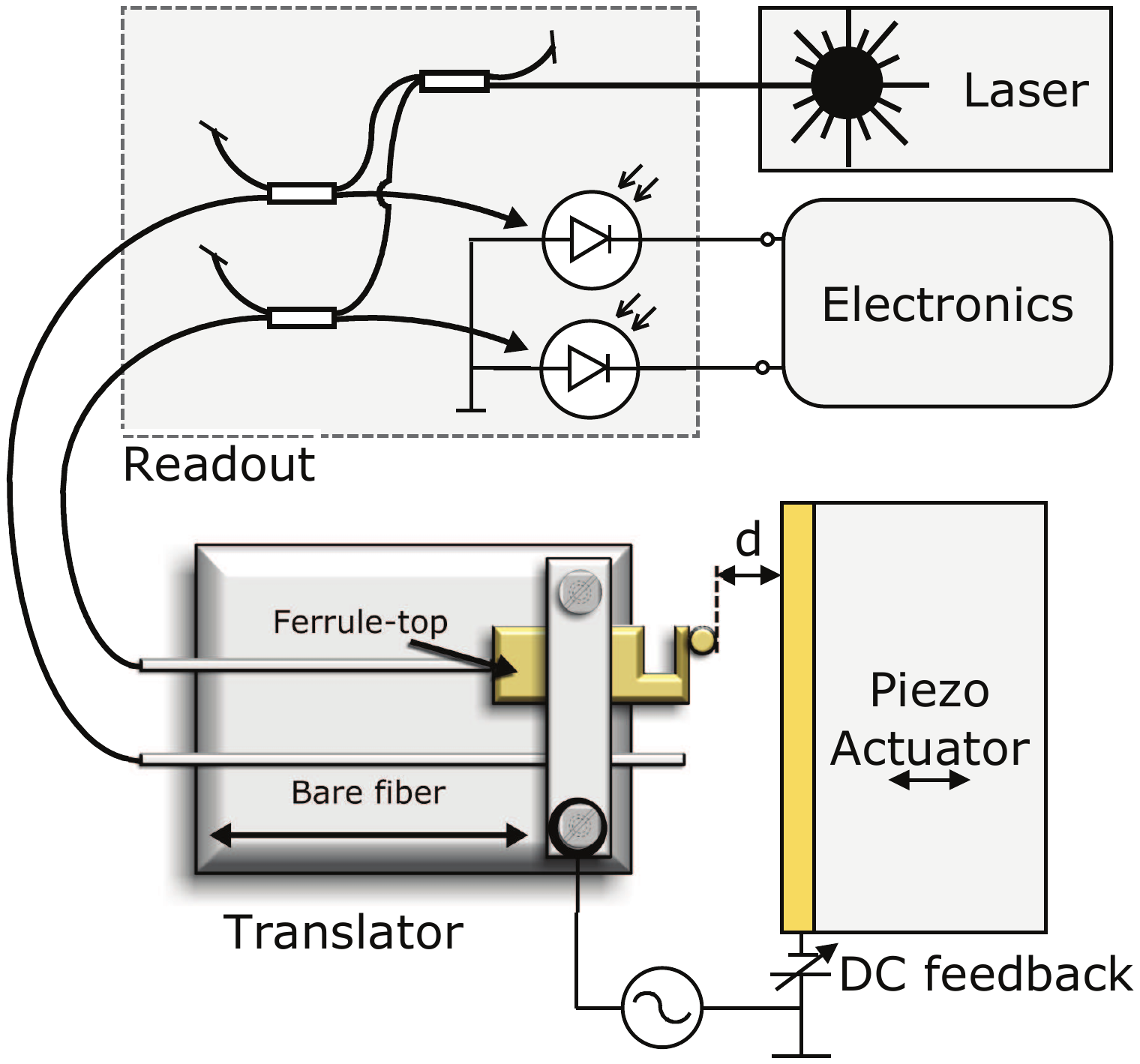}
\label{fig_setup}
\end{center}
\end{figure}

The ferrule-top device is anchored on top of a manual translation stage, just in front of a gold coated sapphire plate that is attached to a piezoelectric stage (see Fig.~\ref{fig_setup}). The manual manipulator allows a first coarse approach of the sensor towards the plate, while the piezoelectric stage is used for the actual scanning during the force-vs-distance measurements. The translational stage also hosts a bare cleaved optical fiber, parallel to the ferrule-top sensor, that is used to measure movements of the piezoelectric stage. The setup is fixed to a block of aluminum that is kept at fixed temperature by means of four resistors controlled via a feedback circuit. To reduce acoustic and seismic coupling to the environment, the whole instrument is mounted on a silicone pad inside an anechoic chamber on top of a marble table equipped with passive vibration damping blocks.

To simultaneously measure the deflection of the ferrule-top cantilever and the motion of the piezoelectric stage, we built two fiber optic interferometers that are fed with the same laser source (Thorlabs Pro800 chassis with a WDM tunable laser module (1552.48~nm to 1554.18~nm)) (see Fig.~\ref{fig_setup}). The laser light is split by a 50/50 optical fiber coupler into two forward branches. In both forward branches, the light is then split again by 90/10 couplers and sent towards the ferrule-top cantilever and the bare cleaved fiber. For the ferrule-top sensor, the light is reflected by the fiber-to-air, air-to-glue, and glue-to-gold interfaces. The amount of light traveling backwards into the fiber is given by
\begin{equation}
W(d_{gap})=W_{0} \left[ 1 + V \cos \left( \frac{4 \pi d_{gap}}{\lambda} + \phi_{0}\right) \right]
\label{eq:interf}
\end{equation}
where $d_{gap}$ is the distance between the fiber end and the cantilever, $W_{0}$ is the mid-point interference signal, $V$ is the fringe visibility, $\lambda$ is the laser wavelength, and $\phi_{0}$ is a phase shift that only depends on the geometry of the cantilever~\cite{Gruca:2010p1677}. This reflected light travels back into the fiber and is split again by the coupler, which sends part of the signal onto a photodetector (Thorlabs PDA10CS). Reading the current generated on the photodetector, which is proportional to $W(d_{gap})$, one can measure changes in $d_{gap}$ (see eq. \ref{eq:interf}) and, thus, the external forces that have produced those changes. The other branch of the double interferometer works identically to the ferrule-top branch, except that the reflected signal is composed of the reflections from the fiber-to-air interface and from the gold mirror, allowing one to measure the relative position of the piezoelectric stage.

From Eq.~\ref{eq:interf} it is clear that it is convenient to operate the force sensor in its quadrature point, where the readout is most sensitive and linear in deflection~\cite{RUGAR:1989p11}. For this reason, before each experiment, we first coarsely bring $d_{gap}$ close to quadrature by adjusting the temperature set-point of the setup, which induces differential thermal expansions on the different parts of the ferrule-top device. We then use the tunable laser wavelength to precisely tune $\lambda$ to the quadrature point\footnote{The 1.7~nm wavelength variation spanned by our laser source alone is not always sufficient to adapt the laser wavelength to the actual length of the fiber-to-cantilever gap.}.

Casimir force measurements are performed following a method similar to that described in~\cite{deMan:2009p99,deMan:2009p104}, which allows one to simultaneously calibrate the instrument, counterbias the electrostatic potential difference that exists between the sphere and the plate, and measure the gradient of the Casimir force as a function of separation.

In a nutshell, while slowly changing the separation between the sphere and the plate by means of the piezoelectric stage, we supply an AC voltage to the sphere with frequency $\omega_{1}$ much smaller than the resonance frequency of the force sensor. This AC voltage gives rise to an electrostatic force that makes the cantilever oscillate. The mechanical oscillation has one component at $\omega_1$ and one component at $2\omega_1$. The $\omega_1$ component drives a negative feedback loop that compensates for the contact potential difference that exists between the sphere and the plate, while the $2\omega_1$ component allows one to calibrate the instrument and to measure the separation between the interacting surfaces. On top of the electrostatic force modulation, we add a small oscillatory motion to the piezoelectric stage at a frequency $\omega_{2}$ that lies somewhere between $\omega_1$ and $2\omega_1$. From the in-phase motion of the cantilever at $\omega_2$, we can finally measure the gradient of the force between the sphere and the plate.

For the details of the experimental method, we refer the reader to~\cite{deMan:2009p99,deMan:2009p104}. It is however important to stress that, contrary to the piezoelectric stage of the setup presented in~\cite{deMan:2009p99,deMan:2009p104}, the one used in this experiment is driven via an open loop circuit and is not equipped with any internal calibration sensor. For this reason, we have implemented a slightly different method to determine the separation between the two surfaces. To explain this new approach, we first note that the electrostatic force generated by the AC voltage is equal to:
\begin{equation}
\frac{F_{e}}{R}=\frac{\varepsilon_0 \pi \left(V_{AC}\cos(\omega_1 t)+V_0\right)^{2}}{d}
\label{eq:elecforce}
\end{equation}
where $\varepsilon_0$ is the permittivity of air, $R$ is the radius of the sphere, and $V_0$ is the residual potential difference. Therefore, the mechanical oscillation induced by the electrostatic force on the force sensor at $2\omega_1$ gives rise to a $2 \omega_{1}$ signal on the photodiode of the interferometer that scales like $S_{2\omega_1} \propto V_{AC}^2/d$. The proportionality constant can be measured by looking at the output signal of the bare fiber interferometer. We know in fact that, when the bare fiber interferometer signal has moved through exactly one interference fringe, the plate has moved for exactly $\lambda / 2$. Once the proportionality constant $\beta$ is known, one can extract $d$ from $d=\beta \cdot V_{AC}^2/S_{2\omega_1}$.

\section{Results and discussion}

The sensor used for the data presented below was a 3.4~mm long, 200~$\mu$m wide, 40~$\mu$m thick ferrule-top cantilever (resulting in an expected spring constant of $\approx 2$ N/m) with $\approx 100$ $\mu$m ferrule-to-cantilever gap (see the scanning electron microscope image of Fig.~\ref{fig_fab}). The resonance frequency was measured independently, and resulted to be equal to 2.7~kHz, with a $Q$ factor of 42.

\begin{figure}[t]
\begin{center}
\includegraphics[width=16cm]{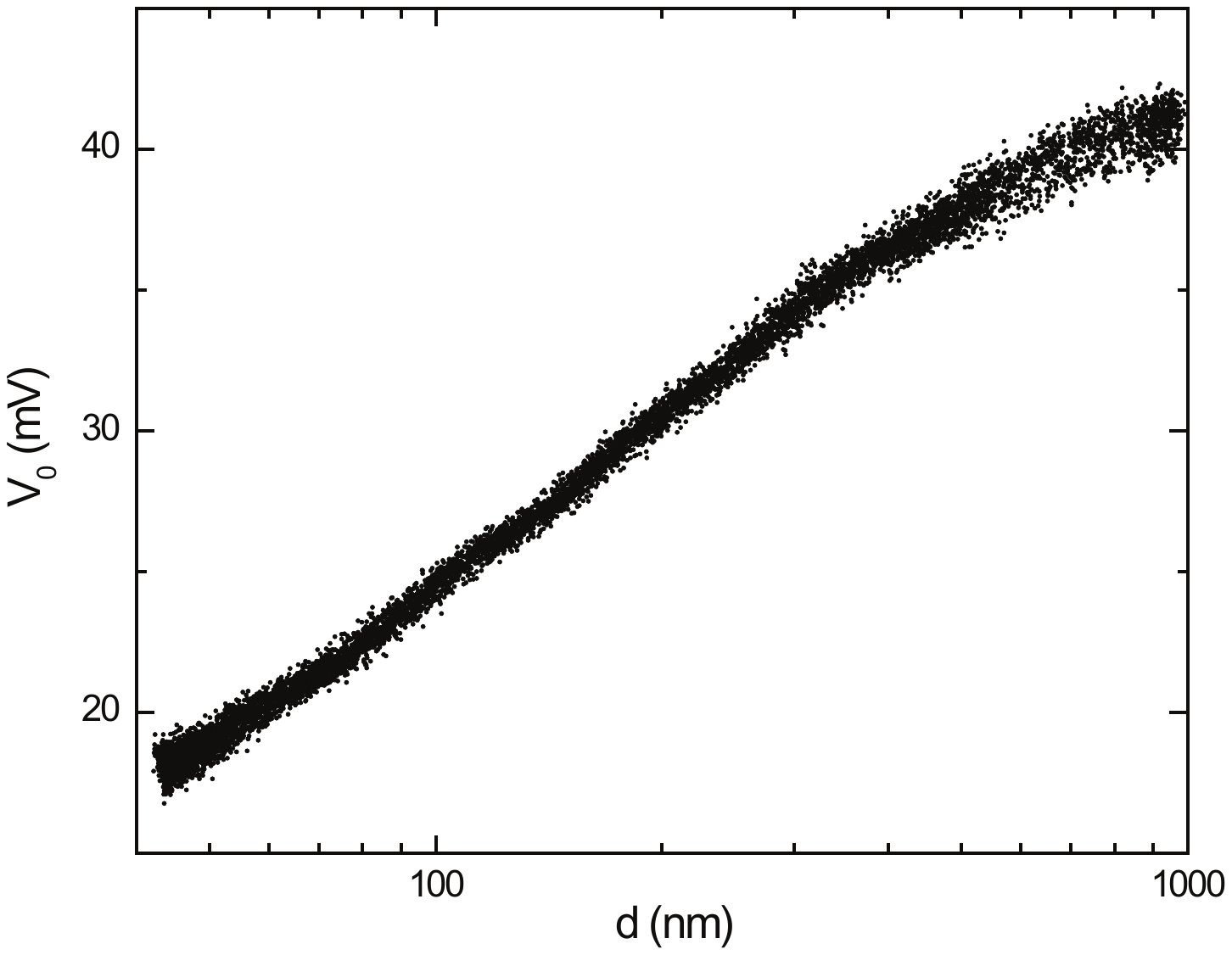}
\label{fig_v0}
\end{center}
\end{figure}

\begin{figure}[t]
\begin{center}
\includegraphics[width=16cm]{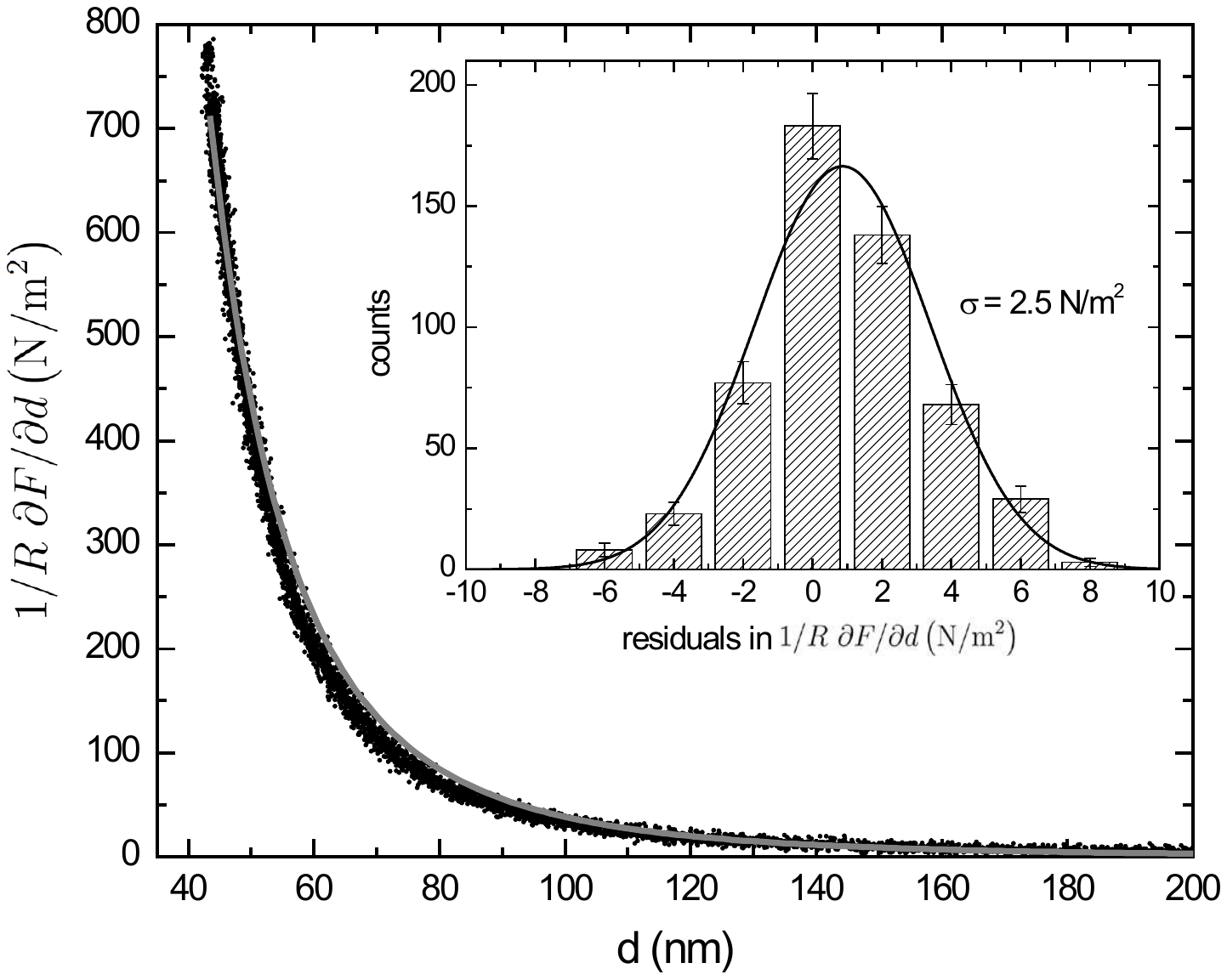}
\label{fig_result}
\end{center}
\end{figure}

In Fig.~\ref{fig_v0} and Fig.~\ref{fig_result} we show the results of a typical measurement run.
Data were gathered during 10 consecutive back-and-forth scans. Each scan had a duration of 1000 s and a stroke of 1 $\mu$m spanned by applying a driving voltage to the piezoelectric stage of the form $V_{\mathrm{PZT}} \propto 1-\left|t/\tau_{s}-1\right|^{3}$, with $\tau_{s}=500$~seconds. The frequency of the AC voltage was set to $\omega_1 = 72$ Hz. Its amplitude was continuously adjusted during the scan to keep the rms of the $2\omega_1$ electrostatic force component equal to roughly 230~pN at all separations (see~\cite{deMan:2009p99,deMan:2009p104}). The oscillation frequency of the piezoelectric stage was set to $\omega_2 = 119$ Hz with 7.2~nm amplitude. Signals at $2\omega_1$ and $\omega_2$ were demodulated with two lock-in amplifiers equipped with a 24dB low pass filter with RC time of 200~ms and 100~ms, respectively. To avoid mixing of the Casimir signal with that induced by the hydrodynamic force due to the air in the gap~\cite{deMan:2009p99}, the phase of the $\omega_2$ lock-in amplifier was aligned with the phase of the oscillatory motion by going to contact, where the plate and the cantilever move synchronously. This procedure was performed only once before starting the measurement run.

Fig.~\ref{fig_v0} shows the potential difference $V_{0}$ needed to minimize the electrostatic interaction between the sphere and the plate as a function of separation $d$. The observed spread in the data is due to measurement noise and not to a time-related drift. It is clear that the data loosely follow a behavior like $a \log d + b$, as observed before in \cite{deMan:2009p104}, \cite{Kim:2009p643}, and \cite{deMan:2010p1654}. This dependence is not yet fully understood.

Fig.~\ref{fig_result} shows the Casimir force gradient as a function of separation. The data were obtained by subtracting from the original data an electrostatic contribution that arises from the calibration procedure~\cite{deMan:2009p99}. This contribution, which scales like $1/d$, can be accurately calculated from the value of $S_{2\omega_{1}}$. In our experiment, this correction ranged from 15~N/m$^{2}$ at 200~nm up to 70~N/m$^{2}$ at 45~nm. The grey line in the graph represents the theoretical Casimir force as computed from the Lifshitz equation, where we have assumed that the dielectric function of the gold surfaces can be obtained by combining the tabulated data of reference~\cite{palig} with the Drude term described in~\cite{lam}, and where we have neglected surface roughness corrections. The theoretical result should thus not be taken too rigorously. It is known, in fact, that gold layers deposited with different methods may have different optical properties, which can lead to significant differences in the resulting Casimir force~\cite{pala}. Furthermore, surface roughness corrections can be as high as several tens of percent at the closest separations. A more refined calculation of the expected force is however outside our scope. The goal of this paper, in fact, is not to improve the accuracy in the comparison between theory and experiment, but to prove that ferrule-top cantilevers can be successfully used to obtain precise (i.e., low noise, small statistical error in force gradient) Casimir force measurements.

It is thus now important to discuss the statistical error in the Casimir force gradient. The inset of Fig.~\ref{fig_result} shows a histogram of the residuals of all the Casimir force data collected in the separation range between 160 nm and 200 nm. The standard deviation is equal to 2.5 N/m$^2$. For comparison, our state-of-the-art atomic force microscope for Casimir force measurements is currently capable of achieving a standard deviation of 1.75 N/m$^2$~\cite{PRAito} with an $\omega_2$ oscillation amplitude a factor of 2 lower but a ten times higher integration time.

\section{Conclusions}

We have presented a ferrule-top sensor for Casimir force experiments. The sensor is based on a monolithic miniaturized cantilever that is coupled to a remote readout via optical fibers. We have demonstrated that the setup provides measurements of the Casimir force between a sphere and a plate by means of a dynamic detection scheme. The sensor can be easily fabricated with cost effective techniques, allowing frequent substitution of the probe in systematic experiments. Furthermore, it adapts well to utilization in harsh environments, such as low temperatures, vacuum, and liquids. Similar ferrule-top devices can of course be used to investigate other long range interaction mechanisms as well. Ferrule-top technology can thus be considered as a new tool to explore phenomena that are of relevance in the future development of MEMS and NEMS.

\section{Acknowledgements}

This project was supported by the European Research Council under the European Community's Seventh Framework Programme (FP7/2007-2013)/ERC grant agreement number 201739, and by the Netherlands Organisation for Scientific Research (NWO) under the Innovational Research Incentives Scheme \textit{Vernieuwingsimpuls} VIDI-680-47-209. The authors acknowledge useful discussions within the community supported by the ESF Research Network CASIMIR.

\pagebreak

\newpage

Fig 1: Fabrication steps followed to manufacture a ferrule-top cantilever for Casimir force measurements. The building block is a pierced 2.5~mm $\times$ 2.5~mm $\times$ 7~mm rectangular ferrule made out of borosilicate glass. The ferrule is machined in the form of a rectangular cantilever, which is then equipped with a spherical bead. An optical fiber slid through the central hole and glued to the ferrule allows detection of cantilever deflections by means of interferometric techniques. The bottom figure is a composition of six scanning electron microscope images showing the device used in the experiment described in the paper.

Fig 2: Sketch of the experimental setup used to measure the Casimir force between a plate and a sphere attached to a ferrule-top cantilever. The ferrule-top cantilever is anchored to a translational stage that allows one to coarsely move the sensor with the sphere close to the plate. The plate is attached to a piezoelectric stage for fine tuning of the separation between the two interacting surfaces. A bare fiber is anchored parallel to the force sensor and is used to measure movements of the piezoelectric actuator via interferometric techniques. An electronic circuit supplies an AC voltage between the sphere and the plate, which allows one to compensate for the residual electrostatic force and calibrate the force sensor. The setup is mounted on an aluminum block kept at fixed temperature inside an anechoic box and isolated from the surroundings with passive vibration dampers (not shown).

Fig 3: Measurement of the residual potential between the interacting surfaces as a function of separation as obtained during 10 consecutive scans.

Fig 4: Dots: Measurements of the gradient of the Casimir force between the sphere and the plate (normalized to the radius of the sphere) as a function of separation as obtained during 10 consecutive scans. The grey line represents the result expected from theory. Inset: histogram of the residuals of the data between 160 nm and 200 nm.

\newpage

\newpage

\section*{References}

\begin{thebibliography}{10}

\bibitem{Casimir:1948p440}
H.~B.~G. Casimir, Proc. K. Ned. Akad. Wet. {\bf 51},  793  (1948).

\bibitem{Capasso:2007p1738}
F. Capasso, J.~N. Munday, D. Iannuzzi, and H.~B. Chan, IEEE J. Sel. Top.
  Quantum Electron. {\bf 13},  400  (2007).

\bibitem{Iannuzzi:2004p117}
D. Iannuzzi, M. Lisanti, and F. Capasso, Proc. Natl. Acad. Sci. U.S.A. {\bf
  101},  4019  (2004).

\bibitem{Lisanti:2005p3}
M. Lisanti, D. Iannuzzi, and F. Capasso, Proc. Natl. Acad. Sci. U.S.A. {\bf
  102},  11989  (2005).

\bibitem{Chen:2006p89}
F. Chen, G.~L. Klimchitskaya, V.~M. Mostepanenko, and U. Mohideen, Phys. Rev.
  Lett. {\bf 97},  170402  (2006).

\bibitem{Chen:2007p15}
F. Chen, G.~L. Klimchitskaya, V.~M. Mostepanenko, and U. Mohideen, Phys. Rev. B
  {\bf 76},  035338  (2007).

\bibitem{Munday:2009p9}
J.~N. Munday, F. Capasso, and V.~A. Parsegian, Nature {\bf 457},  170  (2009).

\bibitem{deMan:2009p99}
S. de~Man, K. Heeck, R.~J. Wijngaarden, and D. Iannuzzi, Phys. Rev. Lett. {\bf
  103},  040402  (2009).

\bibitem{Chan:2008p81}
H.~B. Chan {\it et~al.}, Phys. Rev. Lett. {\bf 101},  30401  (2008).

\bibitem{Torricelli:2010p1674}
G. Torricelli {\it et~al.}, Phys. Rev. A {\bf 82},  010101(R)  (2010).

\bibitem{Lee:2001p1810}
S. Lee and W.~M. Sigmund, J. Colloid. Interf. Sci. {\bf 243},  365  (2001).

\bibitem{parsegian}
V.~A. Parsegian, {\em Van der Waals Forces} (Cambridge University Press, New
  York, 2006).

\bibitem{Lamoreaux:1997p31}
S.~K. Lamoreaux, Phys. Rev. Lett. {\bf 78},  5  (1997).

\bibitem{Kim:2009p643}
W.~J. Kim, A.~O. Sushkov, D.~A.~R. Dalvit, and S.~K. Lamoreaux, Phys. Rev.
  Lett. {\bf 103},  60401  (2009).

\bibitem{Masuda:2009p1887}
M. Masuda and M. Sasaki, Phys. Rev. Lett. {\bf 102},  171101  (2009).

\bibitem{Bressi:2002p1735}
G. Bressi, G. Carugno, R. Onofrio, and G. Ruoso, Phys. Rev. Lett. {\bf 88},
  041804  (2002).

\bibitem{Chan:2001p73}
H.~B. Chan, V. Aksyuk, R.~N. Kleiman, D.~J. Bishop, and F. Capasso, Science
  {\bf 291},  1941  (2001).

\bibitem{Chan:2001p36}
H.~B. Chan, V. Aksyuk, R.~N. Kleiman, D.~J. Bishop, and F. Capasso, Phys. Rev.
  Lett. {\bf 87},  211801  (2001).

\bibitem{Decca:2003p32}
R.~S. Decca, D. Lopez, E. Fischbach, and D.~E. Krause, Phys. Rev. Lett. {\bf
  91},  50402  (2003).

\bibitem{Mohideen:1998p63}
U. Mohideen and A. Roy, Phys. Rev. Lett. {\bf 81},  4549  (1998).

\bibitem{Jourdan:2009p1405}
G. Jourdan, A. Lambrecht, F. Comin, and J. Chevrier, Europhys. Lett. {\bf 85},
  31001  (2009).

\bibitem{deMan:2010p1670}
S. de~Man, K. Heeck, K. Smith, R.~J. Wijngaarden, and D. Iannuzzi, Int. J. Mod.
  Phys. A {\bf 25},  2231  (2010).

\bibitem{Iannuzzi:2006p97}
D. Iannuzzi, S. Deladi, V.~J. Gadgil, R.~G.~P. Sanders, H. Schreuders, and
  M.~C. Elwenspoek, Appl. Phys. Lett. {\bf 88},  053501  (2006).

\bibitem{bao}
M. Bao and H. Yang, Sensors Act. A {\bf 136}, 3 (2007).

\bibitem{Deladi:2006p1739}
S. Deladi, D. Iannuzzi, V.~J. Gadgil, H. Schreuders, and M.~C. Elwenspoek, J.
  Micromech. Microeng. {\bf 16},  886  (2006).

\bibitem{Gruca:2010p1677}
G. Gruca, S. de~Man, M. Slaman, and J.~H. Rector, Meas. Sci. Technol. {\bf 21},
   094033  (2010).

\bibitem{RUGAR:1989p11}
D. Rugar, H.~J. Mamin, and P. Guethner, Appl. Phys. Lett. {\bf 55},  2588
  (1989).

\bibitem{deMan:2009p104}
S. de~Man, K. Heeck, and D. Iannuzzi, Phys. Rev. A {\bf 79},  024102  (2009).

\bibitem{deMan:2010p1654}
S. de~Man, K. Heeck, R.~J. Wijngaarden, and D. Iannuzzi, J. Vac. Sci. Technol.
  B {\bf 28},  C4A25  (2010).

\bibitem{palig}
{\it Handbook of Optical Constants of Solids}, edited by Palik E D 1998 (Academic
Press, New York).

\bibitem{lam}
A. Lambrecht and S. Reynaud, Eur. Phys. J. D {\bf 8}, 309 (2000).

\bibitem{pala}
V. B. Svetovoy, P. J. van Zwol, G. Palasantzas, and J. Th. M. De Hosson, Phys. Rev. {\bf B77}, 035439 (2008).

\bibitem{PRAito}
S. de~Man, K. Heeck, and D. Iannuzzi, Phys. Rev. A {\bf 82}, 062512 (2010).

\end{thebibliography}
\end{document}